# Nuclear Structure Study of Two-Proton Halo-Nucleus $^{17}$Ne

F. H. M. Salih[1], Y. M. I. Perama[1], S. Radiman[1], K. K. Siong[1*]

[1]*School of Applied Physics, Faculty of Science and Technology,*
*Universiti Kebangsaan Malaysia (UKM), 43600 Bangi, Selangor, Malaysia*
[*]khoo@ukm.edu.my

**Abstract**. Theoretical investigation of two-proton halo-nucleus $^{17}$Ne has revealed that the valence protons are more likely to be positioned in the d-state than the s-state. In this study, this finding is clarified by calculation of the binding energy; it is found that the theoretical values for the d-state are closer to the experimental values, in contrast with those obtained for the s-state. The three-body model and MATLAB software are utilised to obtain theoretical values for the three-body-model $^{17}$Ne binding energy and matter radius. $^{17}$Ne has halo properties of a weakly bound valence proton, a binding energy of less than 1 MeV, and a large matter radius. The core deformation parameter has zero and negative values; thus, the $^{17}$Ne core exhibits both spherical and oblate shapes depending on the binding energy of the three-body system. This suggests $^{17}$Ne has two-proton halo.

## 1. Introduction

Since the discovery of halo nuclei in the 1980s, nuclear physics has evolved to yield more notable findings in this field [1]. Halo nuclei are weakly bound exotic nuclei that exhibit unusual states in which the protons and neutrons venture beyond the nuclear drip line [2]. With varied mass distribution, the proton and neutron halo extends far from the nuclear core. One of the first discovered neutron halo nuclei is $^{11}$Li, which was observed in 1985 in experiment at the Lawrence Berkeley Laboratory [1]. Tanihata and colleagues conducted further Berkeley experiments a decade later, proving that the interaction cross sections of $^{6}$He and $^{11}$Li are larger than those for conventional matter radii [1]. This finding opposes the laws of classical physics, as the particles are drifting away from the core nuclear force. According to Hansen and Jonson [3], the halo effect explains why these nuclei are larger than usual, as there is a low binding energy between the core and valence nucleons. An extended halo density is formed, with the core of the $^{11}$Li nucleus ($^{9}$Li) and a di-neutron being weakly bound together [1].

Several techniques can be used to study the detailed structures of exotic nuclei. One of the approach involves application of radioactive nuclear beams for cases in which neutron halos are discovered near the neutron drip line. The halos have been found to exhibit a long tail-like appearance, because of the neutron density distribution in the loosely bound nucleus [2]. In previous studies, halo structures having properties such as low binding energies from the breakup channel, large dissociation cross sections with heavy targets, giant dipole resonance (GDR) for the low-energy halo state, and root mean square (rms) radii larger than the core have been identified [4].

Experimental investigations have shown that neutron halo nuclei such as $^6$He, $^{11}$Li, $^{14}$Be, and $^{17}$B exhibit weaker binding between the core and two valence neutrons compared to non-halo nuclei. This weak binding is indicated by the two-neutron separation energy ($S_{2n}$), and $S_{2n}$ values of −0.973 ± 0.04, −0.369 ± 0.65, −1.34 ± 0.14, and −1.39 ± 0.11 MeV are required to separate the two neutrons and the nucleus core for $^6$He, $^{11}$Li, $^{14}$Be, and $^{17}$B, respectively [5-7]. Thus, $S_{2n}$ is less than 2 MeV [8,9]. As regards the matter radius, the $^6$He (2.50 ± 0.05 fm), $^{11}$Li (3.27 ± 0.24 fm), $^{14}$Be (3.10 ± 0.15 fm), and $^{17}$B (4.10 ± 0.46 fm) nuclei have extraordinarily large radii compared to the matter radii of their cores [10-13]. These two characteristics also apply to proton halo nuclei. For example, $^{27}$S has a two-proton halo and is weakly bound, with a separation energy of 1.4 MeV, along with a large matter radius of 3.07 fm. Meanwhile, as regards one-proton halo nuclei, $^{27}$P has a separation energy of 0.87 MeV with a 3.22-fm matter radius [14,15]. Based on the experimental data, these nuclei can be considered as weakly bound three-body systems composed of a stable core and two nucleons.

The halo-nucleus di-proton $^{17}$Ne has been studied both experimentally and theoretically. The interaction cross section of $^{17}$Ne has been measured in experiment and found to be larger than those of other Ne isotopes. Moreover, the two-proton separation energy has been measured to be −0.933 MeV [16,17]. Fortune and Sherr have suggested that the valence proton is in the d-orbital based on consideration of the Coulomb energy [18]. This has been further reinforced by Millener, who studied beta decay asymmetry on the $^{17}$Ne nucleus and reported that the valence proton is at the d-wave [19].

The $^{17}$Ne nucleus consists of a three-body system ($^{15}$O + p + p), which is also called a 'Borromean structure,' and the two valence protons are weakly bound to the core. A theoretical model is needed to estimate the relative contributions of the s- and d-orbital proton occupancies to the interaction cross section. From few-body Glauber model analysis, it has been suggested that the valence protons of the $^{17}$Ne nucleus occupy the s-state orbital [20]. A three-body model has also been used to determine the structure of $^{17}$Ne, indicating that the s-state and d-state is equally occupied by the valence protons [21,22]. However, analyses based on the Coulomb mass shift have suggested that the d-state is dominantly occupied by the valence protons [23]. Hartree-Fock calculations have also yielded the same results as those given by the Coulomb mass-shift analysis [24].

In the present study, we theoretically investigate the nuclear structure of $^{17}$Ne. We used same approach as neutron halo study like Jacobi coordinates to indicate the bound states of the two valence protons in order to produce the wave function and the total Hamiltonian of three-body [25]. In addition, the total Hamiltonian of the three-body system, there is a new addition was applied in this study which is the Coulomb interaction as Coulomb effect in core-proton and proton-proton interaction. The total Hamiltonian of the three-body system and the rms matter radius operator are used to calculate the theoretical binding energy of the three-body system and the rms matter radius of $^{17}$Ne, respectively. We primarily focus on the core deformation to determine the binding energy of the three-body system and the rms matter radius. Consequently, we use experimental data to obtain the core deformation parameter. In addition, we study the correlation of the s- and d-bound states, for which the Coulomb interaction in included in the theoretical calculation. MATLAB is used for all calculations.

## 2. Methodology

A halo nucleus has two nucleons bound to the core. Thus, this is a three-body system (core + proton + proton) identical to the Borromean configuration. In this study, the core is studied using the shell model while the entire system is analysed based on a cluster model. Fig. 1 shows the Jacobi coordinates used to model the system in this study.

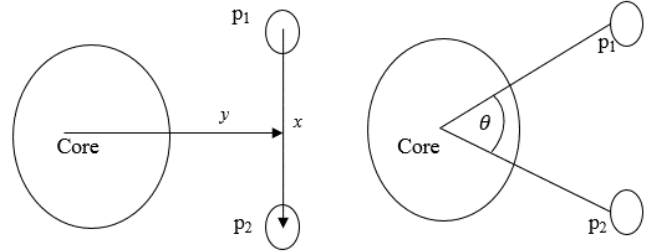

**Fig 1** Overview of three-body system configuration.

The Jacobi coordinates $(x, y)$ are incorporated in the wave function of the system:
$$\psi^{JM}(x, y, \xi) = \phi_{core}(\xi_{core})\psi(x, y) \qquad (1)$$

The information on the radius, angle, and spin of the two protons is expressed in the wave function $\psi$. The relative motion in the three-body system is described using hyperspherical coordinates incorporating masses $A_i$, where $i$ = 1, 2, and 3. Here, $(x, y)$ indicate the distance between each pair of particles, $\vec{r}_{jk}$, and the distance between the center of mass of a particle pair and the third particle. Thus,

$$x = \sqrt{A_j}\vec{r}_{jk} = \sqrt{\frac{A_j A_k}{A_j + A_k}}\vec{r}_{jk} \text{ and } y = \sqrt{A_{(ij)i}}\vec{r}_{(jk)i} = \sqrt{\frac{(A_j A_k)A_i}{A_j + A_k + A_i}}\vec{r}_{(jk)i}. \qquad (2)$$

Further, in this study, $(x, y)$ were converted to hyperspherical coordinates of hyper-radius $\rho$ and hyper-angle $\theta$ according to the relations

$$\rho^2 = x^2 + y^2 \text{ and } \theta = \arctan(\frac{x}{y}) \tag{3}$$

where the hypershperical expansions of the radial wave function of the three-body system and the angular wave function are, respectively [25],

$$R_n(\rho) = \frac{\rho^{\frac{5}{2}}}{\rho_0^3}\sqrt{\frac{n!}{(n+5)!}} L_{nlang}^5(z)\exp(\frac{-z}{2}) \tag{4}$$

$$\psi_k^{l_x l_y} = N_k^{l_x l_y}(\sin\theta)^{l_x}(\cos\theta)^{l_y} P_n^{l_x+\frac{1}{2},l_y+\frac{1}{2}}(\cos 2\theta). \tag{5}$$

In this definition, $\rho = \sqrt{j(j+1)}$, $z = \rho/\rho_0$, $\rho_0 = \sqrt{m_j(m_j+1)}$, $j = l_x + 1$, and $m_j = -j, j+1, \ldots, j$, where $j$ is the total angular momentum and $l_x$ is the angular momentum of a valence proton. The associated Laguerre polynomials of order $nlang = 0,1,2,3,\ldots$ are defined as $L_{nlang}^5$, while the normalisation coefficient is defined as $N_k^{l_x l_y}$ and $k = l_x + l_y + 2n$ is the hyperangular momentum quantum number with $n = 0,1,2,\ldots$. Further, $P_n^{l_x+\frac{1}{2},l_y+\frac{1}{2}}(\cos 2\theta)$ in Equation (5) is the Jacobi polynomial, where $n = l_x + 1$. The wave function of a valance nucleon (proton) is

$$\psi_{n,k}^{l_x l_y}(\rho,\theta) = R_n(\rho)\psi_k^{l_x l_y}(\theta). \tag{6}$$

The details about the formalism of the hyperspherical harmonics method are presented in Refs. 25 and 26. Then, the Hamiltonian $\hat{H}$ of the three-body system is [27,28]

$$\hat{H} = \hat{T} + \hat{h}_{core}(\xi) + V_{core-p1}(r_{core-p1},\xi) + V_{core-p2}(r_{core-p2},\xi) + V_{p-p}(r_{p-p}) \tag{7}$$

In this equation, the kinetic energy is $\hat{T} = \hat{T}_x + \hat{T}_y$ and $\hat{h}_{core}(\xi)$ is intrinsic Hamiltonian of the core, which depends on the internal variables $\xi$. Further, $V_{core-p_i}$ and $V_{p-p}$ represent two-body interactions for all pairs of interacting bodies, while the Coulomb interaction (Vc) is added between the core-proton and proton-proton interaction. The deformed Wood-Saxon potential are considered in the overall potential, where:

$$V_{core-p}(r_{core-p_i},\vec{\xi}) = \frac{-V_0}{[1+\exp(\frac{r_{core-p_i}-R(\theta,\emptyset)}{a})]} + \frac{-\hbar^2}{m^2c^2}(2l.s)\frac{V_{so}}{4r_{core-p}}\frac{d}{dr}[1+\exp(\frac{r_{core-p_i}-R_{s.o}}{a})]^{-1} +$$

$$Vc(r_{core-p_i}). \tag{8}$$

The interaction for proton-proton is defined as [27,28]

$$V_{p-p}(r_{p-p_i}) = \frac{-\hbar^2}{m^2c^2}(2l.s)\frac{V_{so}}{4r_{p-p}}\frac{d}{dr}[1+\exp(\frac{r_{core-p_i}-R_{s.o}}{a})]^{-1} + Vc(r_{p-p_i}), \tag{9}$$

and the Coulomb interaction is defined as

$$V_c(r_{i-j}) = \frac{e^2}{4\pi\varepsilon_0}[\frac{1}{|\vec{r_1}-\vec{r_2}|}]. \tag{10}$$

In Equation (8), $R(\theta,\emptyset) = R_0[1 + \beta_2 Y_{20}(\theta,\emptyset)]$ is the radius that depends on the core quadrupole deformation parameter $\beta_2$, where $\theta$ and $\emptyset$ are spherical angles in the core rest frame,

$R_0 = 1.25 A^{1/3}$, and $R_{s.o} = R$. Further, $l$ is the orbital momentum operator between the core and a proton, $s$ is the proton spin operator, and $m$ is the pion mass.

In this study, for the core-proton interaction, we used a deformed Woods-Saxon potential, spherical spin-orbit term and introduced Coulomb interaction in Equation (8). The nucleon-nucleon interaction was approximated using spin orbit interaction included with Coulomb interaction.

The rotational model was applied to the core structure; hence, the core was modelled as having axially symmetric-rotor deformation. Therefore, the radius of this deformed core was expanded in spherical harmonics [25].

For an A-body system, the operator for the average squared distance of a nucleon from the position of the total centre of mass is determined using the following expressions:

$$\vec{r}_{CM} = \frac{1}{A}\sum_{i=1}^{A} \vec{r}_i, \tag{12}$$

$$r_m^2 = \frac{1}{A}\sum_{i=1}^{A}(\vec{r}_i - \vec{r}_{CM})^2, \tag{13}$$

$$<r_m^2>^{\frac{1}{2}} = \frac{1}{A}[A_{core}<r_m^2(core)> + <\rho^2>], \tag{14}$$

where $\vec{r}_i$ is the position of the $i$th nucleon, $\vec{r}_{CM}$ is the centre of mass, and $<r_m^2>^{1/2}$ is the rms matter radius.

The above derivations have yielded a number of important equations. In particular, Equation (6) is a wave function for a two-valence-proton system and Equation (1) is the total wave function of the three-body system. This wave function is the result of multiplication of the internal wave functions of the cluster (those of the core and protons). The total wave function includes correlation of the valence proton and core movements, where Jacobi coordinates are used to describe the configuration of the three-body system, as shown in Fig 1. Further, in Equation (4), the total angular momentum ($j$) of the valence proton depends on the core-p radius; hence, approximations of $\rho$ and $\rho_0$ were calculated. Equations (7) and (14) are used to calculate the binding energy of the three-body system and the rms matter radius of the nucleus. The binding energy of the three-body system depends on the Wood-Saxon potential, spin-orbit interaction, and simple Coulomb interaction, as shown in Equation (8).

Throughout the present work, the spin-orbit term was left deformed with a radius of $r_o = 1.25$ fm, where $R_{S.O} = R$. The diffuseness was fixed to the standard value of $a = a_{s.o} = 0.65\ fm$. Calculations were performed for the full range of $\beta_2$, i.e., $-0.7$ to $0.7$. The properties of $^{17}$Ne were calculated upon variation of $\beta_2$ with fixed spin orbit depth $V_{s.o}$ of $-15$ MeV, and central Wood-Saxon depth $V_o$ of $-74$ MeV [29]. The value of $\beta_2$ indicates one of three nucleus shapes: oblate (negative $\beta_2$), spherical ($\beta_2 = 0$), or prolate (positive $\beta_2$).

All the concepts discussed above were applied to $^{17}$Ne to observe the relationship between the undeformed and deformed core and the binding energy and rms matter radius. Table 1 lists the parameter values used in the computational process, where MATLAB version R2013b calculations were conducted to produce theoretical values approximate to the experimental value for the binding energy of $^{17}$Ne.

**Table 1.** Parameter values used in calculations.

| $l_x$ | $l_y$ | n | K | $\rho$ | $\rho_0$ | $r_0$ (fm) | a (fm) | $a_{S.O}$ (fm) | $V_0$ (MeV) | $V_{S.O}$ (MeV) |
|---|---|---|---|---|---|---|---|---|---|---|
| 0 | 0 | 1 | 2 | 0.866 | 0.866 | 1.25 | 0.65 | 0.65 | -74 | -15 |
| 1 | 1 | 2 | 6 | 0.866 | 1.936 | 1.25 | 0.65 | 0.65 | -74 | -15 |
| - | - | - | - | 1.936 | - | 1.25 | 0.65 | - | - | - |
| 2 | 2 | 3 | 10 | 0.866 | 2.958 | 1.25 | 0.65 | - | - | - |
| 2 | 2 | 3 | 10 | 1.936 | - | 1.25 | - | - | - | - |
| 2 | - | - | - | 2.958 | - | 1.25 | - | - | - | - |

## 3. Results and Discussion

The $^{17}$Ne nucleus consists of a core ($^{15}$O) and valence nucleon (proton) with an experimentally measured separation energy of approximately −0.933 MeV [29]. In this study, the binding energy of the $^{17}$Ne nucleus for the s- and d-states was calculated based on a three-body system. Table 2 lists the theoretical binding energies for the three-body system for the s- and d-states. The parameter $\beta_2 = 0.00$ was selected based on the shell model. The $^{17}$Ne core is comprised of $^{15}$O having $Z = 8$; thus, it also has a magic number (closed shell) and the core does not experience deformation (i.e., it has a spherical shape). Fig. 2 shows the relationship between the binding energy of the three-body system and the core deformation for the s- and d-states. Tables 2 and 3 list the theoretical values of the binding energy and $\beta_2$ for the $^{17}$Ne three-body system.

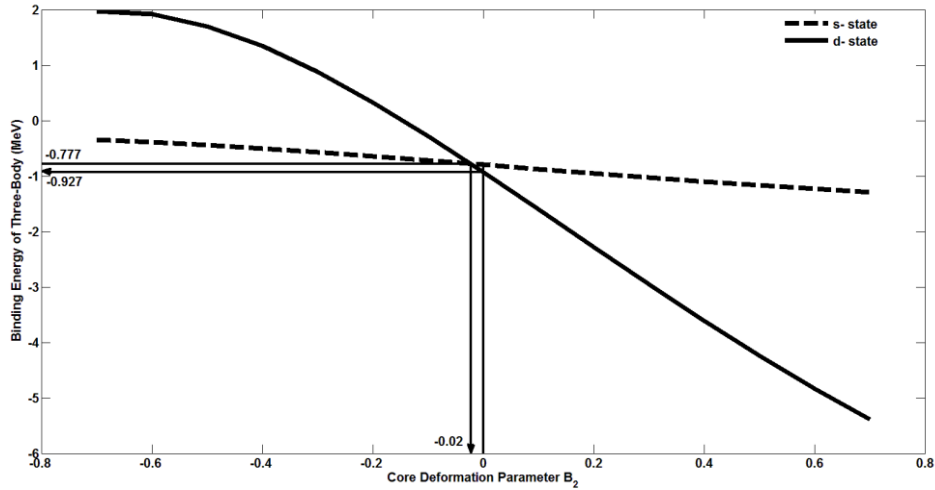

**Fig 2.** Binding energies of $^{17}Ne$ three-body system as functions of core deformation for $2s_{1/2}$ and $1d_{5/2}$ states.

**Table 2.** Theoretical binding energies for $^{17}Ne$ three-body system.

| Nucleus | Deformation parameter, $\beta_2$ | Binding energy of three-body system (MeV) | |
|---|---|---|---|
| | | $2s_{1/2}$ state | $1d_{5/2}$ state |
| Oxygen-15 | 0.00 | | |
| Neon-17 | | −0.795 | −0.927 |

**Table 3.** Theoretical values of core deformation parameter $\beta_2$ of $^{17}Ne$.

| Nucleus | Binding energy (MeV) from previous experiment | $\beta_2$ | |
|---|---|---|---|
| | | $2s_{1/2}$ state | $1d_{5/2}$ state |
| Oxygen-15 | | 0.18 | nearly zero |
| Neon-17 | −0.933 [29] | | |

Based on Fig 2 and 3, the theoretical values of the binding energy of the three-body system are −0.795 ($2s_{1/2}$ state) and −0.927 MeV ($1d_{5/2}$ state) at $\beta_2 = 0.00$. Thus, the binding energies for both states are below 1 MeV for a potential depth $V_0$ of −74 MeV. Based on the low binding energy of the three-body system, the valence protons of $^{17}$Ne have the potential to occupy both the s- and d-states. Halo nuclei have low angular momenta, with $l = 0$ and 1 for the s- and p-states, respectively. However, the $^{17}$Ne valence proton in the d-state has high angular momentum. The binding energies calculated at d-state were higher compared to s-state. The binding energy of three-body in this study show that the value for the d-state is closer to the experimental value. Hence, it can be concluded that there is a high probability of valence proton occupation at the d-state. This statement is contradicted by the findings of previous studies. That is, Kanungo et al. have found that the interaction cross sections of $^{17}$Ne are larger at the s-state than the d-state [16]. In addition, Tanaka et al. have shown that $^{17}$Ne has a long-tailed density at the s-state [17] and Geithner et al. have stated that $^{17}$Ne has a large charge radius when the valence proton is in the s-state configuration [30]. There was an intersection point between s- and d-state at $\beta_2 = -0.02$, which shared same binding energy of three-body, -0.777 MeV respectively. At this point, the valence proton has probability to have a mixing configuration between s- and d-state or sd state. This statement is strengthened by Fortune and Sherr [23]

Through normalisation using experimental data for the binding energy, $\beta_2 = 0.18$ ($2s_{1/2}$ state) and $\beta_2 = 0.00$ ($1d_{5/2}$ state) have been obtained. These two values indicate that $^{17}$Ne exhibits a prolate and spherical shape. The values of core deformation parameter $\beta_2 = -0.02$ and 0.00 were used to determine the values of rms matter radii of $^{17}$Ne. The binding energy result agree with the experimental data.

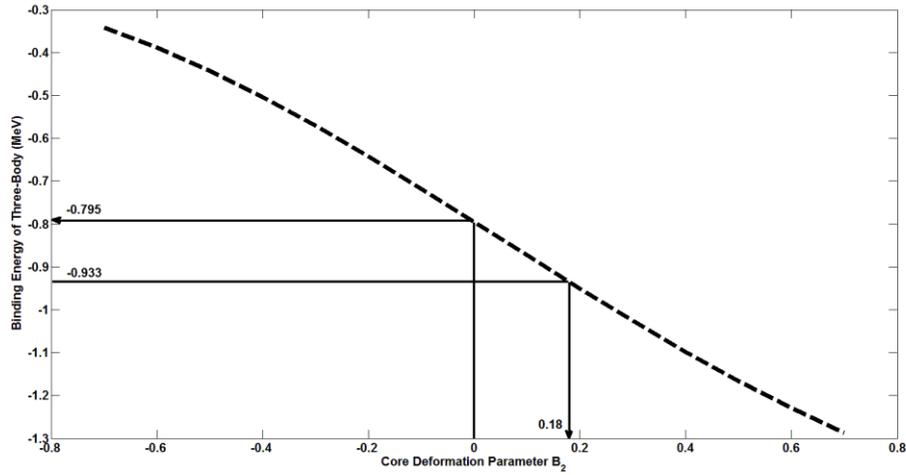

**Fig 3.** Binding energies of $^{17}Ne$ three-body system as functions of core deformation at $2s_{1/2}$.

**Table 4**. Theoretical values of binding energy of $^{17}Ne$ three-body system.

| Nucleus | $\beta_2$ | Binding energy of three-body system (MeV) | |
|---|---|---|---|
| | | $2s_{1/2}$ state | $1d_{5/2}$ state |
| Oxygen-15 | -0.02 | | |
| Neon-17 | | −0.777 | −0.777 |

Another aspect indicating halo nucleus formation is extension to a large density distribution or expansion of the nuclear radius. This is due to the quantum mechanical tunnelling from the nuclear volume, which is induced by short-range force combined with low binding energy, causing valence nucleon tunnelling into the classical region [8,9]. In this study, experimental data on the binding energy of the three-body system and the matter radius of $^{17}$Ne were used to calculate the deformation of the $^{15}$O core. Based on Figs. 2 and 3, values core deformation $\beta_2$ of $^{17}$Ne were obtained, as presented in Table 6. Selection of core deformation $\beta_2$ was based on the dominant occupancy of proton at d-state. These values were used to determine theoretical values of the matter radius of $^{17}$Ne, which are shown in Fig. 4 and Table 6.

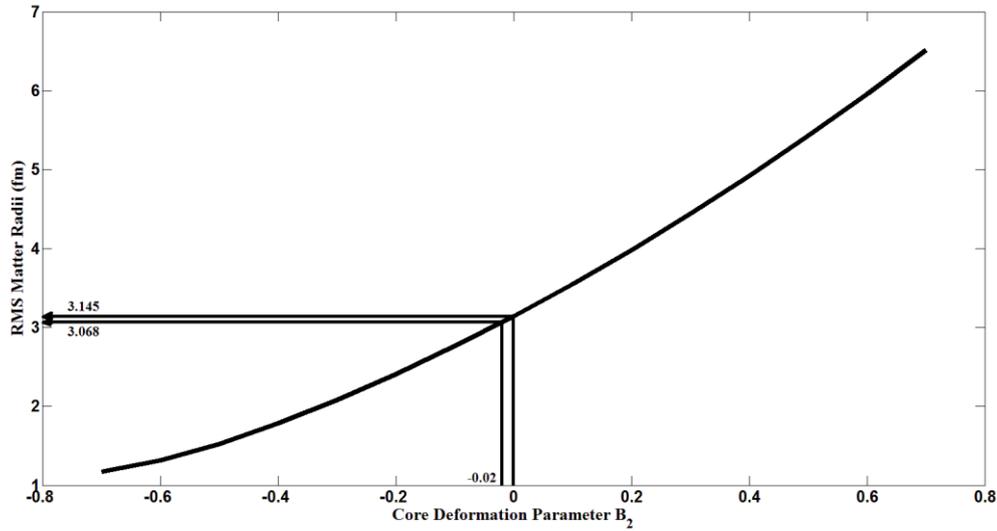

**Fig 4.** Rms matter radius of $^{17}Ne$ as function of core deformation.

**Table 5.** Theoretical values of $^{17}Ne$ $\beta_2$ and rms matter radii.

| Nucleus | $\beta_2$ | Binding energy (MeV) from previous experiment | Radius (fm) |
|---|---|---|---|
| Oxygen-15 | +0.00 −0.02 | | |
| Neon-17 | | −0.933 [32] | 3.145 3.068 |

The matter radii of $^{17}$Ne were obtained through normalisation using the obtained $\beta_2$ values, as listed in Table 6. Previously, Ozawa et al. reported that the matter radius of $^{17}$Ne is 2.75 ± 0.07 fm [31]. Angeli and Marinova have since provided an experimentally obtained and updated matter radius value for $^{17}$Ne of 3.041 ± 0.0188 fm [32]. The theoretical values obtained in the present study are 3.145 and 3.068 fm for $\beta_2$ = 0.000 and -0.02, respectively, which are closer to the updated experimental data. The core radius of $^{17}$Ne ($^{15}$O) is reported to be 2.44 ± 0.04 fm [33], and addition of the two valence protons to the $^{15}$O nucleus yields a matter radius of 3.0413 ± 0.0088 fm [32]. This shows that a matter radius abnormality occurs in $^{17}$Ne. Finally, based on the liquid drop model mass formula and the rms matter radius formula newly adjusted by Royer and Rousseau, the $^{17}$Ne rms matter radius is 3.15 fm [34]. The experimental data reported by Angeli and Marinova [32] can be used as a reference, because the values reported by Royer and Rousseau [34] and those obtained in the present study are closer to each other. Based on Table 7, $^{17}$Ne has a larger rms matter radius than other isobaric nuclei.

**Table 6.** Experimentally obtained matter radii of $^{17}C$, $^{17}N$, $^{17}O$, $^{17}N$, and $^{17}Ne$ and values obtained in present work.

| Nucleus | Rms matter radius (fm) |
|---|---|
| $^{17}$C | 2.72 ± 0.03 [32] |
| $^{17}$N | 2.48 ± 0.05 [32] |
| $^{17}$O | 2.6932 ± 0.0075 [33] |
| $^{17}$F | 2.54 ± 0.08 [32] |
| $^{17}$Ne | 3.0413 ± 0.0088 [33] |
|  | 3.145 (present work) |
|  | 3.068 (present work) |

Based on the findings of this study, the core of $^{17}$Ne is important for analysis of the halo characteristics, particularly the binding energy of the three-body system and the matter radius. Using the results presented in Figs. 2 and 3, the core deformation parameters were determined through normalisation (i.e., experimental values were used to calculate theoretical values). Negative (slightly deformed) and zero core deformation parameters were obtained, as reported in Table 5. These results indicate that the $^{17}$Ne core may have an oblate or spherical shape depending on the binding energy of the three-body system and the state possibly occupied by the valence proton.

## 4. Conclusion

In conclusion, we have used experimental data to investigate the nuclear structure of $^{17}$Ne, focusing on the core deformation parameter. Through a normalisation approach, theoretical values were obtained for this parameter. Further, based on the findings of this study, the three-body model can be applied to determine the structure of the $^{17}$Ne halo nucleus. Specifically, the binding energy of the $^{17}$Ne three-body system was found to be less than 1 MeV. The calculated matter radii were 3.145 and 3.068 fm according to the different theoretical values of the core deformation parameter; these values were closest to the experimental data.

The theoretical values obtained in this study indicate that this model meet an agreement with the experimental data. According to the calculated results, valence protons dominantly occupy the d-state rather than the s-state and allows the sd mixing configuration. This three-body model can be used to determine the characteristics of other proton halo nuclei in future studies.

## Acknowledgment

The authors would like to thank the Ministry of Higher Education of Malaysia for financial support provided through the fundamental research grant scheme (grant number: FRGS/2/2013/ST02/UKM/02/1).